\documentclass[12pt]{elsart}

\usepackage{psfig,latexsym}

\begin{document}

\begin{frontmatter}
\title{The role of step edge diffusion in epitaxial crystal growth}
\thanks[me]{Corresponding author. FAX:+49-931-888-4604; e-mail: schinzer@physik.uni-wuerzburg.de}
\author{S. Schinzer\thanksref{me}},
\author{M. Kinne},
\author{M. Biehl},
\and
\author{W. Kinzel}
\address{Institut f\"ur Theoretische Physik,
         Julius-Maximilians-Universit\"at W\"urzburg,
         Am Hubland, D-97074 W\"urzburg, Germany}
\date{\today}

\begin{abstract}
The role of step edge diffusion (SED) in epitaxial growth is
investigated.  To this end we revisit and extend a recently introduced
simple cubic solid-on-solid model, which exhibits the formation and
coarsening of pyramid or mound like structures. By comparing the
limiting cases of absent, very fast (significant), and slow SED we
demonstrate how the details of this process control both the shape of
the emerging structures as well as the scaling behavior. 
We find a sharp transition from significant SED to intermediate values
of SED, and a continuous one for vanishing SED.
We argue that one should be able to control these features of the surface
in experiments by variation of the flux and substrate temperature.
\end{abstract}

\end{frontmatter}

\section{Introduction}
The investigation of non-equilibrium growth processes has
attracted considerable attention in the past decades. 
Despite the lack of a general
theory as in thermal equilibrium systems, significant progress has
been achieved using the concept of self-affine  growing surfaces
(see \cite{bar95} for an introduction and overview). 

Molecular beam epitaxy (MBE) is a standard technique to obtain high 
quality crystals needed for e.g.\ semiconductor devices. In MBE 
particles are evaporated from an oven, cross a vacuum
chamber and are deposited on a substrate. The
particles diffuse on the growing surface and a relatively low incoming 
flux allows for the formation of  nearly perfect crystals.
Much effort has been devoted to a theoretical understanding of MBE 
growth and the various morphologies and 
scaling behaviors observed in experiments, see e.g.\
\cite{kp95} for an overview of experimental investigations and 
\cite{kru97b} for a recent review of theoretical approaches.
Several simple models have been proposed to describe specific aspects
of MBE \cite{wv90,dg92}, but no unified scaling picture has yet
emerged.  Many of these models fail to explain the frequently observed
slope selection, i.e.\ the formation of pyramids or mounds on the
growing surface which show a constant inclination (independent of
time) \cite{tkw95,sps95,eff94,zw97,joh94,ojl95,oeg98}.  In terms of a
continuum description slope selection has been suggested to be due to
the balance of upward and downward currents on the surface
\cite{sp94,rk97}. Only recently, processes similar to step edge
diffusion have been considered in the study of continuum equations
\cite{sie99,gdn99}. 
So-called full diffusion models of MBE frequently take into account
step edge diffusion \cite{sv95,fb97,bfk98,ssw97,af96a}. There,
microscopic energies are used to parameterize activated jump rates of
the particles. However, simulations of these models are very time
consuming which prevents a precise determination of the scaling laws.
In the (commonly used) simplest models SED is not included
explicitly\footnote{The diffusion along a step edge or the detachment
away from a step have the same energy barrier.} \cite{sv95}.  Whereas
the effect of SED on submonolayer growth has been adressed by several
authors \cite{fb97,bfk98,ssw97}, its influence on mound formation and
coarsening has not been investigated \cite{sv95,af96a}.

Recently, we have introduced a simple cubic solid-on-solid model of
epitaxial growth \cite{bks98} in which slope selection emerges from
the competition of two mechanisms:
\begin{enumerate}
\item[(A)] When a particle arrives at the surface its residual
  momentum perpendicular to the substrate enables it to move to a site
  of lower height, if such a site is available within a neighborhood of
  radius $R_i$.
\item[(B)] Otherwise, the particle diffuses on a terrace of the
  surface until it binds to an upward step or meets another diffusing
  particle. This latter {\sl nucleation} is taken into account in a
  simple manner by means of a maximum diffusion length $\ell_D$.
  Furthermore, a large Ehrlich-Schwoebel barrier rules out downward
  moves to lower terraces.  Hence, such downward moves are only possible
  upon arrival through process (A).
\end{enumerate}
The incorporation (A) of particles constitutes a downward current
(downward funneling or transient diffusion \cite{est90,yhp98}) whereas
the Schwoebel effect (B) yields an effective upward current
\cite{eh66,ss66}.  Both currents are slope dependent and their
cancellation gives rise to the formation of mounds with a constant
inclination as shown in \cite{bks98}.  At later stages, mounds grow
and merge until a single one remains which occupies the entire
substrate.

In this article we demonstrate that the features of the growing
surface depend crucially on another mechanism, the diffusion of
particles along step edges (SED):
\begin{enumerate}
\item[(C)] A particle which is  incorporated (A) or reaches an edge by  
  diffusion (B) can hereupon move randomly along the edge until it finds 
  an  additional in-plane neighbor at a kink site. 
  In analogy to process (B), SED is also limited by a maximum
  diffusion length $\mbox{$\ell_{\mbox{\scriptsize SED}}$}$. 
\end{enumerate}
We want to mention, that a similar model has been studied by Bartelt and Evans
\cite{be95}. However, they achieved smooth step edges not by means of
SED. Rather, they artificially enforced square based shaped islands by
explicitly rearranging them after the attachment of particles without
treating the intersection of islands in a realistic way.

In this paper we investigate the following  limiting cases for
$\ell_{\mbox{\scriptsize SED}}$ (note that all length scales will be
given in lattice constants):
\begin{itemize}
\item[-] {\sl significant} SED ({\em i.e.} $\ell_{\mbox{\scriptsize
      SED}}$ of the order of the system size $L$) in Sec.~\ref{sSED},
\item[-] {\sl vanishing} SED ($\ell_{\mbox{\scriptsize SED}}\!=\!1$,
  Sec.~\ref{vSED}) which is shown to be comparable to growth with an
  additional Ehrlich--Schwoebel barrier for SED (Sec.~\ref{EsSED})
  which we will refer to as {\sl restricted} SED,
\item[-] and no SED ($\ell_{\mbox{\scriptsize SED}}\!=\!0$) in
  Sec.~\ref{noSED}.
\end{itemize}
In addition, the crossover for intermediate values of
$\ell_{\mbox{\scriptsize SED}}$ is investigated in
Sec.~\ref{crossover}. Concluding remarks are given in
Sec.~\ref{conc}. 

\section{Model description}                 \label{model}
We consider the deposition of particles on  
 a simple square lattice and assume that a solid-on-solid
restriction is fulfilled,  i.e.\  the surface can be  
described by an integer data array of heights $h(x,y)$.  The 
discussion is restricted to models with an incorporation radius
$R_i=1$ in process (A). Greater values should only change the 
average terrace width of the resulting mounds. 
A simple argument for this was already given in \cite{bks98}: 
 the frequency of downward moves (A) on a given terrace of width
$\ell_T$  is proportional
to $R_i$. If $\ell_D > \ell_T$ the upward current is proportional
 to $\ell_T - R_i$. Only for $\ell_T = 2 R_i$ these currents cancel and
 slope selection is achieved. The diffusion on terraces (B) is 
simulated explicitly as a random walk which is restricted to sites
of equal height. The assumption of an infinite Ehrlich--Schwoebel
barrier is not unrealistic. Values of 0.2 eV have been reported for
metal homoepitaxy \cite{sto94,sh95}. At room temperature this implies
that diffusion is by a factor 2000 faster than jumps over step
edges. Clearly, smaller Ehrlich--Schwoebel barriers increase the
downward current and lead to larger terrace widths \cite{sk99}.  

Starting from a flat substrate in MBE, the initial number of mounds 
can be controlled by the flux of the deposited particles. In
our model, this effect (deriving from the interplay between many
particles) is described in a simplistic manner: If a particle has not
found an upward step edge after $n_D$ moves it is treated as if it had 
met another particle and formed an immobile nucleus. The diffusion 
length $\ell_D$ is thus related to the distance of islands in the 
sub-monolayer regime and corresponds to the mean free path in a density 
of diffusing particles. On compact non-fractal terraces 
a maximum number of diffusion steps $n_D$ corresponds to
 $\ell_D =\sqrt{n_D}$.  Theories \cite{vpt92} and simulations 
\cite{jkw97} predict in this case that  $\ell_D \propto 
\left(D/F\right)^{1/6}$ with 
the diffusion constant $D$ and the flux $F$ of arriving
particles. In our model neither the temperature nor the flux can be
varied explicitly but enters indirectly through $\ell_D$.
In all simulations presented here we have set
$\ell_D=15$, which merely fixes the initial number of
mounds on the substrate \cite{bks98}. 
Since SED is an effectively one-dimensional process
we can replace it by a deterministic search for the nearest
kink site  within a distance $\pm \mbox{$\ell_{\mbox{\scriptsize SED}}$}$
along the step edge.  To begin with, we assume that no
further restriction
applies to the SED; the presence of a barrier at step edge corners will 
be discussed later. Again, the characteristic length 
$\mbox{$\ell_{\mbox{\scriptsize SED}}$}$ 
can be related to the reduced flux $f$ (per unit length) and the 
corresponding diffusion
constant $\delta$:~ $\mbox{$\ell_{\mbox{\scriptsize SED}}$}
 \propto \left(\delta / f \right)^{1/4}$. 

Only after the particle has reached its
final position  a new particle is placed  randomly on the surface.   
This very efficient algorithm allows for the simulation of large
systems over several decades in time which is well beyond the
capabilities of conventional full diffusion models.
As our model does not include the possibility of desorption,
time can be expressed in terms of {\sl monolayers\/}, i.e.\ 
the total number of deposited particles divided by the 
system size $L^2$.

\section{Limiting cases}
In all models considered here, an initial fast formation of mounds is
observed. In the presence of SED (\ref{sSED},\ref{vSED},\ref{EsSED})
the slope remains unchanged in the subsequent coarsening
process. Hence the ratio of the lateral mound size and their height
remains constant. Only in the absence of SED (\ref{noSED}) we observe
time dependent slopes.

\subsection{Significant SED}                \label{sSED}
\begin{figure}[t]
\begin{center}
\setlength{\unitlength}{1.0cm}
\begin{picture}(8.2,9.4)(0,0)
  \put(-0.2,5.2){\makebox(4,4){
\psfig{file=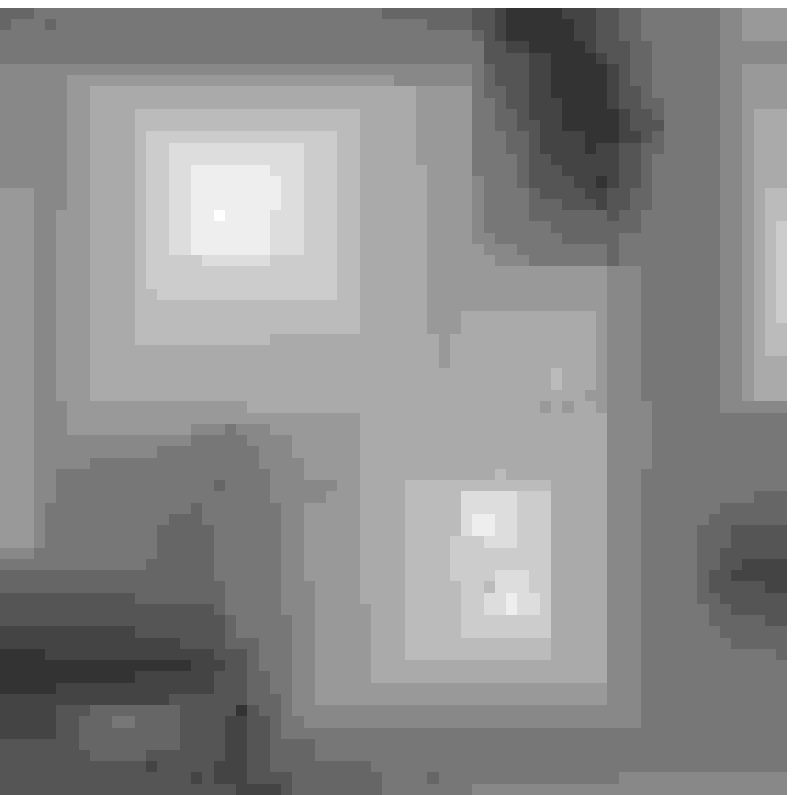,width=4.1cm}}}
\put(4.2,5.2){\makebox(4,4){
\psfig{file=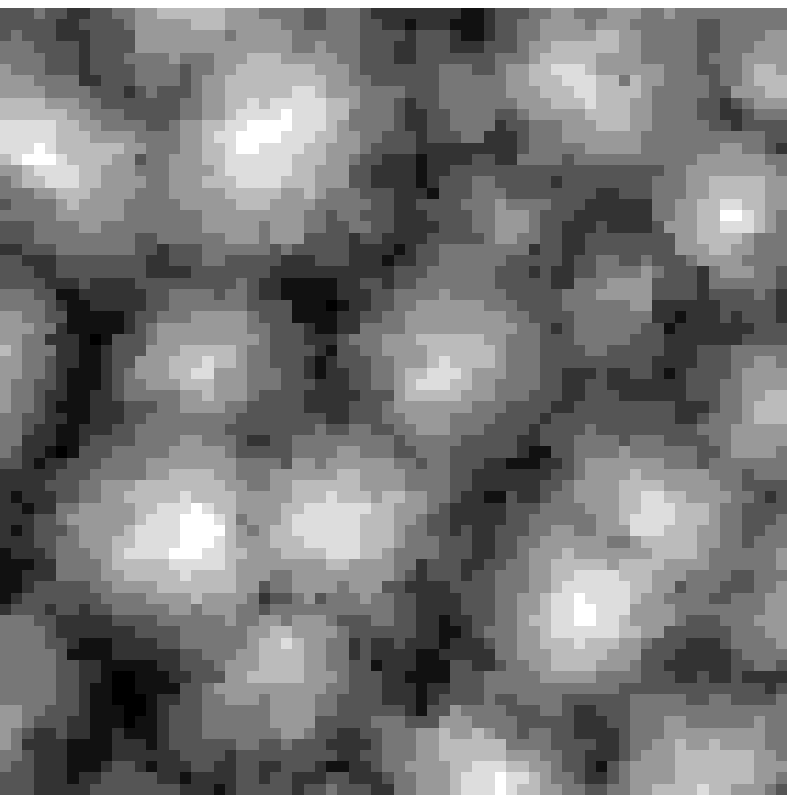,width=4.1cm}}}
  \put(-0.2,0.3){\makebox(4,4){
      \psfig{file=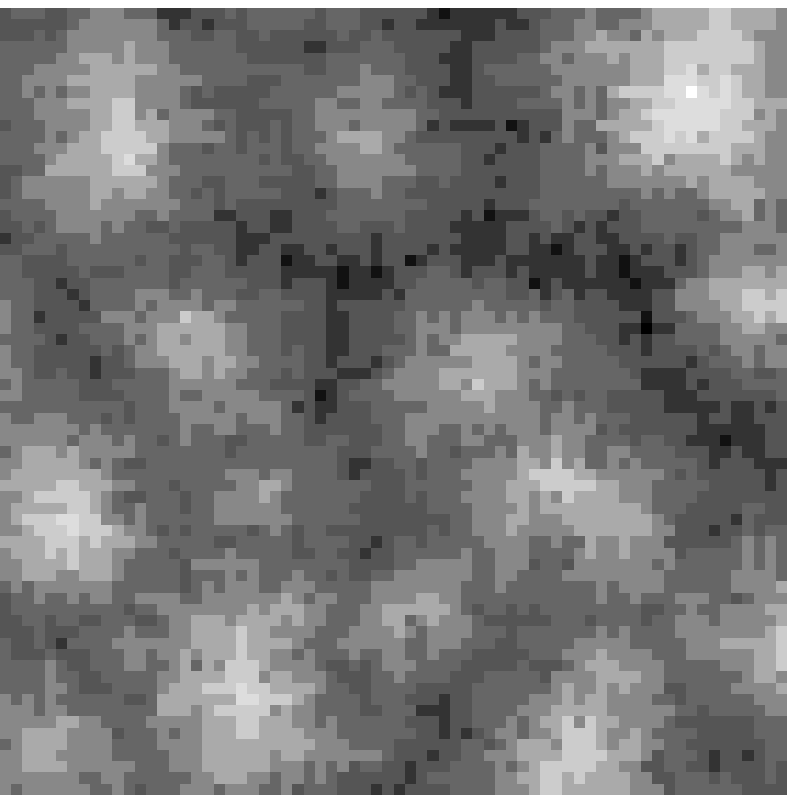,width=4.1cm}}}
  \put(4.22,0.3){\makebox(4,4){
      \psfig{file=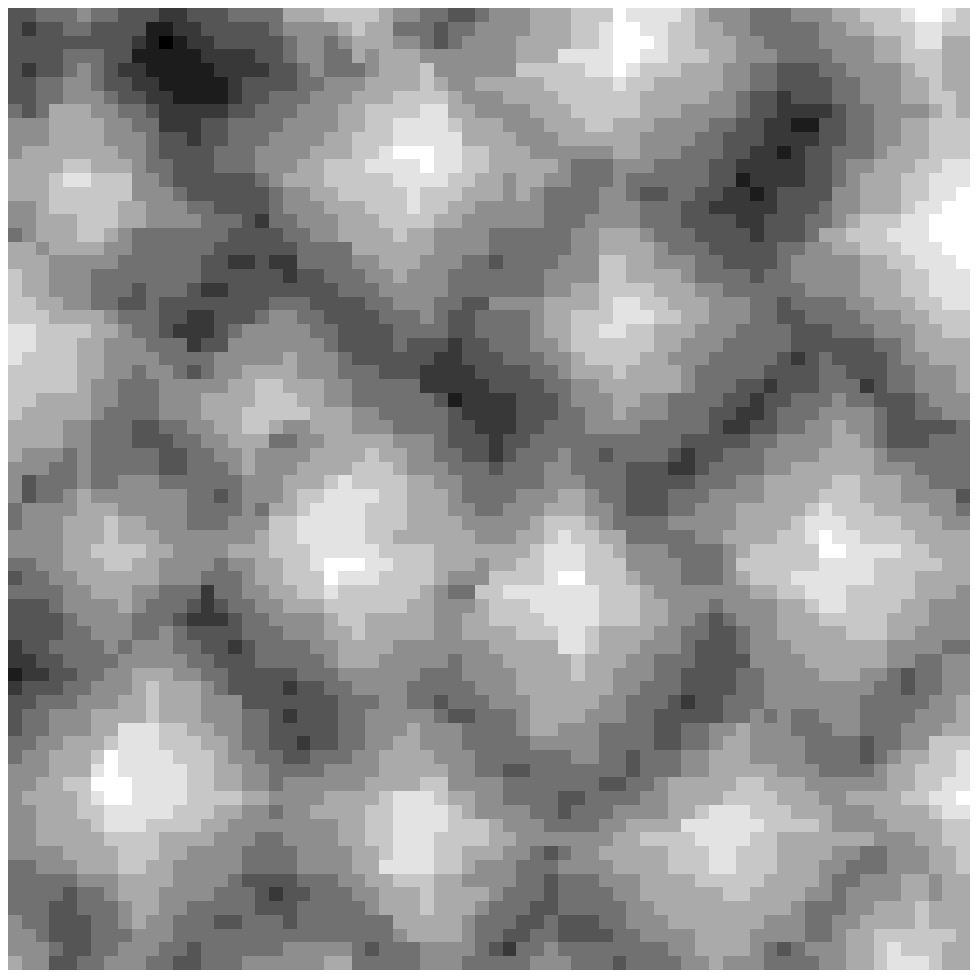,width=4.1cm}}}
  \put(0.0,4.7){\mbox{\bf a)}}
  \put(4.25,4.7){\mbox{\bf b)}}
  \put(0.0,-0.1){\mbox{\bf c)}}
  \put(4.25,-0.1){\mbox{\bf d)}}
\end{picture}                    
\caption{                               \label{snapshots}
  Snapshots of the surfaces after growth of 64 (panel a) or 128 
  monolayers (all others). Windows of size $70\times 70$ sites are shown;
  simulations were performed on a  square lattice with $L=140$,
  $\ell_D = 15$. 
  Pictures correspond  to:  $\mbox{$\ell_{\mbox{\scriptsize
SED}}$}=2L$ (panel a), $\ell_{\mbox{\scriptsize
  SED}}=1$ (b), 
  no SED ($\mbox{$\ell_{\mbox{\scriptsize SED}}$}=0$) (c), 
and SED in the presence of an infinite barrier 
 at corners (d) with $\mbox{$\ell_{\mbox{\scriptsize SED}}$}=2$. }
\end{center}
\end{figure}

In our earlier investigation of the limit of significant SED
($\ell_{\mbox{\scriptsize SED}} = {\mathcal O}(L)$) we observed two
distinct regimes: First, pyramids (their number is of the order of
$L^2/\ell_D^2$) with a well defined slope form and their square shaped
bases reflect the lattice structure, cf.~Fig.~\ref{snapshots}(a).
Then, a coarsening process starts which is characterized by two
scaling exponents.  The surface width
\[w = \sqrt{\left\langle (h(x,y)-\langle h \rangle)^2 \right\rangle}\]
increases according to a power law $w \propto t^\beta$ with a growth
exponent $\beta \approx 0.45$. We observe that $w$ saturates after a
time $t_\times \propto L^{z}$ at a value $w_{sat} \propto L^{\alpha}$
where the dynamic exponent $z$ satisfies $z=\alpha/\beta$. This
corresponds to the scaling picture of Family and Vicsek for self-affine
surfaces \cite{fv85}.  In the saturation state only one pyramid of the
size of the system remains in our model. As a direct consequence of slope
selection the saturation width increases linearly with $L$. Hence the
roughness exponent is $\alpha =1$ (and $z = 1/\beta \approx 2.20$). 

It is intuitively clear that fast SED should result in a very rapid 
coarsening of the structures. Whenever two pyramids merge, as for
instance shown in Fig.\ 1 (a), material is moved efficiently 
along the straight edges towards the region where the base terraces
touch and a large density of kink sites is provided. Hence
fast SED  fills the inner corners at the regions of contact.

\subsection{Vanishing SED}           \label{vSED}
\begin{figure}
\begin{center}
\centerline{    \psfig{file=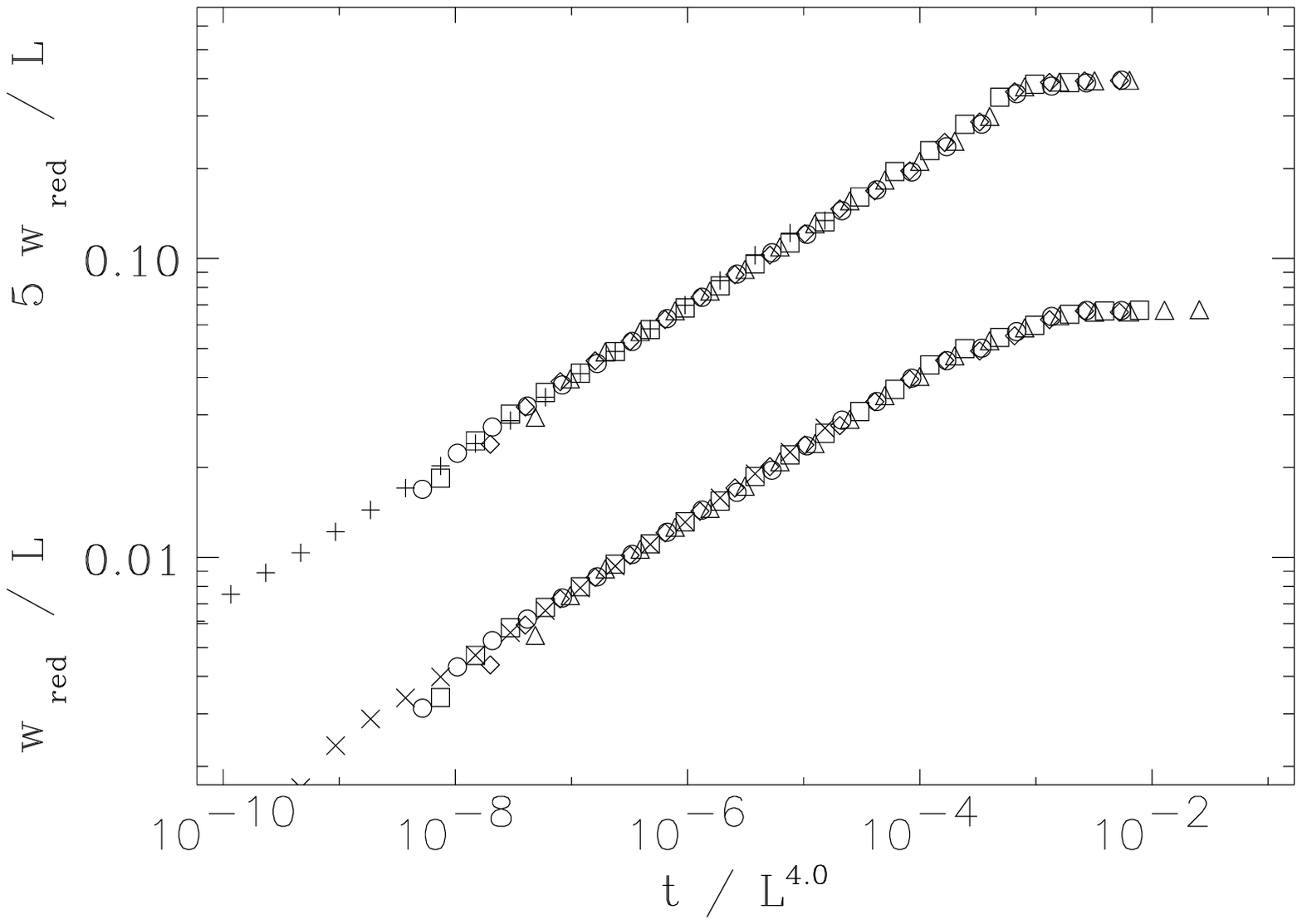,width=8cm}}
\caption{                   \label{w-wi}
 The evolution of the reduced surface width for the model  
 with restricted SED with 
$\mbox{$\ell_{\mbox{\scriptsize SED}}$}=2$  (upper curve, $w_{red}$ 
multiplied by 5) and without a barrier at step edge corners with 
$\mbox{$\ell_{\mbox{\scriptsize SED}}$}=1$ (lower curve). 
Here, the values $w_i = 0.62$ (upper) and $w_i = 0.67$ (lower curve)  
have been used for all system sizes, $L= 80 (\bigtriangleup), 
 100 (\Diamond), 128 (\Box), 140 (\mbox{\Large $\circ$}), 
 256 (\times), 512 (+) $. Averages are over 10 independent 
 simulations, standard error bars would be smaller than the size 
of the symbols.}
\end{center}
\end{figure}

Systems in which the particles are nearly immobile at step edges still
exhibit slope selection with $\alpha =1$, but otherwise a
significantly different scaling behavior is found.  In this section we
will investigate the limiting case $\ell_{\mbox{\scriptsize SED}} =
1$. The extension to larger $\ell_{\mbox{\scriptsize SED}}$ of order
one will be postponed until Sec.~\ref{crossover}. Fig.~\ref{w-wi}
(lower curve) shows the evolution of the surface width with time for
$\ell_{\mbox{\scriptsize SED}}=1$. Here, an intrinsic width $w_i$ has
been taken into account.  Following \cite{bar95} we have plotted $
w_{red} = \sqrt {(w^2 - w_i^2)^2} $ where $w_i$ is an $L$-independent
contribution to the total width. In our model, one source of the
intrinsic width  is clearly the initial slope selection process.
The value of $w_i = 0.67$ is close to the estimate $\ell_D/(2
R_i \sqrt{72}) \approx 0.88$ for perfect square pyramids of size
$\ell_D$ and slope $1/2 R_i$ which are formed within the first $t_0$
monolayers of growth. As $w_i$ is less than one monolayer, it
can be neglected in the long time limit.
Curves for different system sizes
collapse under the scaling assumption for the reduced width $w_{red}$
with $\alpha = 1$ and $z=4.0$ respectively $\beta = 0.25$.

The lack of a rapid transportation of material along edges has two
consequences for the surface morphology.  First, as
expected, the shape of the growing mounds is rounded, as can be
seen in Fig.~\ref{snapshots}(b). No long range smoothening mechanism
favors the orientation of edges along the lattice vectors.
Furthermore, the coarsening is driven mainly by fluctuations of the
particle deposition and hence is much slower than the SED assisted
process.

The observed exponent $z =  4.0$ coincides with the results
of \cite{sp94,rk97} obtained in continuum models
without explicit SED mechanism.  It is also interesting
to note the agreement with a recent hypothesis of
Tang, {\v S}milauer, and Vvedensky \cite{tsv97} that purely noise assisted
coarsening should be characterized by  $ z = 2+d$ for growth
on  $d$-dimensional substrates in the presence of slope
selection. 

A particular difficulty should be noted with respect to the 
observation of the saturation behavior. In the late stages of 
growth only a few mounds are present. The merging of these last 
structures seems to depend crucially on the actual configuration,
i.e.\  their relative position. This was already observed for
the case of fast SED and can lead to a cascade like increase of $w$
close to saturation with {\sl metastable\/} configurations 
persisting over relatively large times.

\subsection{Ehrlich--Schwoebel effect for SED}           \label{EsSED}
Next we discuss the influence of an energy barrier for SED around
corner sites. In analogy to  the Schwoebel-effect one can argue that
moves around the corners of terrace edges should be suppressed. Indeed,
there is experimental and theoretical evidence  
for such an effect, see for instance 
\cite{fb97,ssw97a} and references therein. 

As a clear limiting case we consider the presence of an infinite
barrier at terrace corners, referred to as restricted SED in the
following. Hence, the movement of particles through SED (process C) is
always limited to a straight portion of the edge.  If no additional
binding partner can be reached within a distance $\pm
\mbox{$\ell_{\mbox{\scriptsize SED}}$}$, the particle remains
immobile, hereby increasing the density of kinks. We find that the
favored shape of terraces corresponds to squares which are tilted with
respect to the lattice vectors by $45^o$ ({\sl diamonds}) and offer
the maximum possible number of kink sites.  In Fig.~\ref{snapshots}(d)
a snapshot of an accordingly growing surface is shown.  Pyramids with
diamond shaped bases and a stable slope have emerged in contrast to
the rounded {\sl cones \/} of Fig.~\ref{snapshots}(b). This
observation becomes obvious if one considers an initially square
shaped pyramid base. A particle diffusing at a straight step edge will
be reflected by the corners which leads to an effective flux towards
the center of the step edge. Therefore, extended terraces aligned with
the lattice axis are unstable.

Despite the obvious differences in surface morphology, the scaling
behavior of the systems with and without restriction of SED is
approximately the same for small $\mbox{$\ell_{\mbox{\scriptsize
SED}}$}$.  In Fig.~\ref{w-wi} we compare the evolution of the
corresponding (reduced) width $w_{red}$; the data rescale under the
assumption $ z= 4.0$ and $\beta = 0.25$ in both cases.  Again, the
coarsening is driven mainly by deposition noise and hence it is
relatively slow compared to the model with fast, unrestricted SED.

Note that the actual value of $\mbox{$\ell_{\mbox{\scriptsize SED}}$}$
should be largely irrelevant in restricted SED.  After a
($\mbox{$\ell_{\mbox{\scriptsize SED}}$}$-dependent) transient,
diamond shaped terraces should always emerge, along which particles
can diffuse only a few steps in any case.  We will study the validity
of this simplifying argument in a forthcoming project.

\subsection{Growth without SED}            \label{noSED}
\begin{figure}[t]
\begin{center}
\centerline{  \psfig{file=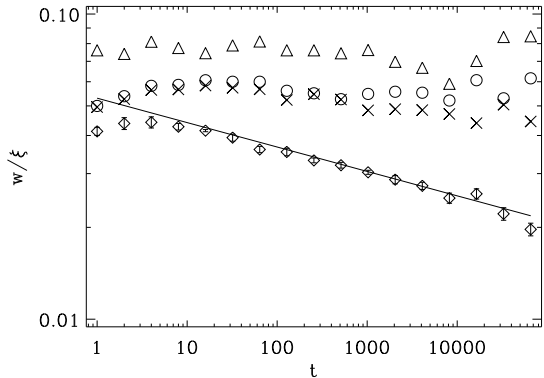,width=8cm}}
\caption{                                   \label{wdxi}
  Temporal evolution of the surface width $w$ divided by the typical
  distance $\xi$ on the surface. The symbol ($\bigtriangleup$) represents
  the model with restricted SED, here $\mbox{$\ell_{\mbox{\scriptsize
  SED}}$}=2$, system size $L=512$, 
  11 independent simulation runs (the last 6 points are only from 3
  runs). Unrestricted SED: ($\times$) $\mbox{$\ell_{\mbox{\scriptsize
        SED}}$}= 1$, $L=256$, 70 runs; (\mbox{\Large $\circ$})
  $\mbox{$\ell_{\mbox{\scriptsize SED}}$}=2$, $L=256$, 70 runs.
  In the absence of SED ($\Diamond$, $L=512$, 44 runs, the last two
  points are from 21 runs, standard error bars are shown) $w/\xi$
  follows a power law, the solid line corresponds to $(\alpha-1)/z=
  -0.08$.  
} 
\end{center}
\end{figure}

Fig.~\ref{snapshots}(c) shows a snapshot of a surface grown in the absence
of SED. As for unrestricted SED with small but non-zero 
$\mbox{$\ell_{\mbox{\scriptsize SED}}$}$,
the emerging structures
are rounded. However, the total lack of smoothening SED leads to 
very rugged terrace edges.  As we will see below, the system
does not select a stable constant slope of the mounds. 
Apparently, the absence of well defined extended terraces makes 
the effective cancellation of downhill and uphill currents impossible. 
Clearly, our simple argument yielding a mean terrace width  $2 R_i$
is not applicable  here. 

The above mentioned difficulty in observing saturation is
particularly pronounced for 
$\mbox{$\ell_{\mbox{\scriptsize SED}}$}=0$. {\sl Metastable\/} surface 
configurations
with very few remaining mounds persist already in fairly small systems and
reduce the usefulness of scaling plots  analogous to Fig.~\ref{w-wi}. 
Hence, we have determined  first $\beta$ from the simulation of large 
systems without attempting to  achieve saturation. We find a value of
$\beta \approx 0.20$, which clearly demonstrates the importance of 
step edge diffusion: the mere presence of SED with
 $\mbox{$\ell_{\mbox{\scriptsize SED}}$}=1$ already 
changes the growth exponent to $\beta = 0.25$ as shown above.  

In order to determine a second scaling exponent we have analysed the
Fourier transform of the growing surfaces. 
For all simulations at all times a pronounced maximum in the Fourier 
spectrum is observed, which can be used
to extract a typical distance $\xi$ on the surface. 
The scaling behavior of $\xi$, which corresponds to the typical mound
size, is given by the dynamic exponent:~ $\xi \propto t^{1/z}$. 
Numerically we obtain a value of $ z < 4 $ which together with $\beta
\approx 0.20 $ implies $ \alpha < 1$ in the absence of SED.  
In Fig.~\ref{wdxi} the quantity $w/\xi$ is plotted which scales 
like \( w/\xi \propto t^{(\alpha -1)/z} \) with time.  The upper three
curves correspond to the system with slow SED 
($\mbox{$\ell_{\mbox{\scriptsize SED}}$}=1,2$) and 
the model with restricted SED (here 
$\mbox{$\ell_{\mbox{\scriptsize SED}}$}=2$).  
In all these cases $w/\xi$ is roughly constant for large
times as expected from
the observed slope selection with $\alpha =1$.  However, 
in the absence of SED the quantity follows a power with an exponent  
$ (\alpha - 1)/z $ in the vicinity of $-0.08$, i.e.\ 
$\alpha \approx 0.7$, which clearly indicates the absence of
slope selection.  Note that these findings do support the hypothesis 
of \cite{tsv97} that for general noise assisted coarsening 
$2\beta + d/z = 1$, which is fullfilled in this case with $\beta=0.2$,
$\alpha = 0.7$, and $z = \alpha/\beta = 3.5$.

The consequences of $\alpha < 1$ for the persistence of mounds in the
limit $t \rightarrow \infty$ remains unclear at this stage, even
though we have realized $t \approx 10^5$. Our main point here,
however, is to demonstrate the absence of slope selection.

\section{Crossover behavior}            \label{crossover}
\begin{figure}[t]
\centerline{  \psfig{file=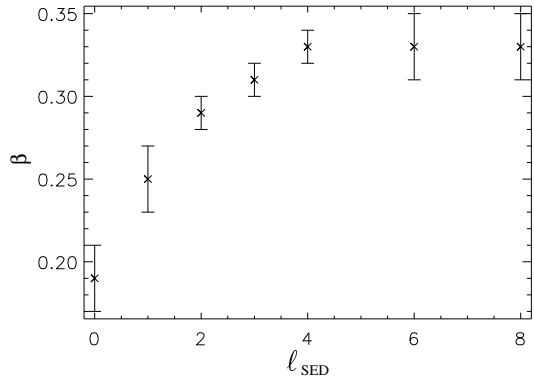,width=8cm}}
\caption{                     \label{lowSED}
  Continuous rise of $\beta$ from 0.25 to 0.33 for small values of
  $\ell_{\mbox{\scriptsize SED}}$.  $2^{15} = 32768$ ML were deposited
  on a system of size $256 \times 256$. The diffusion length was set
  to $\ell_D = 15$. The $w(t)$ curve was averaged over at least 7
  independent simulation runs. Error bars are standard deviations of a
  nonlinear fit to the power law in the range 4 ML -- 32768 ML.
}
\end{figure}

We now turn to the investigation of intermediate values of
$\ell_{\mbox{\scriptsize SED}}$. One could speculate that for any
finite $\ell_{\mbox{\scriptsize SED}} \ll L$ the limiting case of
vanishing SED ({\em i.e.} $\ell_{\mbox{\scriptsize SED}} = 1$) should
be reached as $t \to \infty$.  However, fig.~\ref{lowSED} supports a
value of $\beta = 0.33$ for all $\ell_{\mbox{\scriptsize SED}} \ge
4$. In these simulations we deposited $2^{15} = 32768$ ML but no
indication for a crossover to $\beta = 0.25$ was found. We believe
that the mean terrace width (here: $2 R_i = 2$) 
determines the value of $\ell_{\mbox{\scriptsize SED}}$ at which
$\beta = 0.33$ is reached. Clearly, the lateral fluctuations of the
step edges are related to SED. Higher values of
$\ell_{\mbox{\scriptsize SED}}$ reduce the lateral fluctuations. On
the other hand, if the fluctuations become so large that they sense
the terrace border, the growth behavior should change. We will
investigate the validity of this speculative argument in a forthcoming
project. We should be able to confirm our hypothesis by increasing
$R_i$ or reducing the Ehrlich--Schwoebel barrier, both yielding an
increased mean terrace width.

\begin{figure}
\begin{center}
\centerline{  \psfig{file=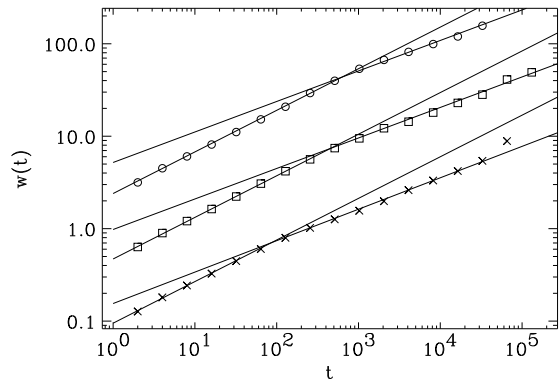,width=8cm}}
\caption{   \label{cross}
Crossover from fast SED to vanishing SED for large, intermediate
values of $\ell_{\mbox{\scriptsize SED}}$. The straight lines
correspond to $w \propto t^{0.33}$ and $w \propto t^{0.45}$. The
simulations were carried out on a $1024 \times 1024$ lattice with
$\ell_D = 15$. The different symbols correspond to
$\ell_{\mbox{\scriptsize SED}}=$ 50 ($\times$), 80 ($\Box$), and 100
({\footnotesize$\bigcirc$}).
}
\end{center}
\end{figure}

Both exponents, $\beta = 0.33$ and $\beta = 0.45$, are relevant values
of the growth exponent as can be most easily identified in
fig.~\ref{cross}. There, we investigate the crossover for models with
relatively high values of $\ell_{\mbox{\scriptsize SED}}$ compared to
the terrace width. Therefore, the simulation of large systems was
necessary and we chose $L=1024$. The time needed for the simulations
was several weeks on a HP9000/715. As a consequence, the data are
calculated from a single simulation run. Nevertheless, our results
support the following picture: Initially, the structures are small
compared to $\ell_{\mbox{\scriptsize SED}}$. Hence, we obtain the
scaling behavior of significant SED with $\beta = 0.45$. When the
typical size $\xi$ of the structures becomes of the order of
$\ell_{\mbox{\scriptsize SED}}$ the SED is no longer significant and
the growth exponent drops down to $\beta = 0.33$.  The time $t_\times$
when the crossover occurs therefore depends on $\ell_{\mbox{\scriptsize
SED}}$ like $t_{\times} \propto \ell_{\mbox{\scriptsize SED}}^z$
(since $\ell_{\mbox{\scriptsize SED}} \approx \xi \propto
t_{\times}^{1/z}$) which is in accordance with the data presented in
fig.~\ref{cross}.  It should be possible to observe the scaling with
$\beta = 0.45$ in experiments if SED is sufficiently fast in the
investigated material. One might speculate that this is the case in
\cite{eff94} where a value $\beta \approx 0.5$ was observed in the
growth of Cu(001) at $T = 200 \mbox{K}$.

\section{Conclusions}                \label{conc}
In summary, our findings show qualitatively and quantitatively that
the features of epitaxial growth processes depend crucially on the
presence of step edge diffusion and its detailed properties. This
offers a possible explanation for the large variety of surface
morphologies and scaling properties observed in experiment and
simulation.  In particular, since the SED--constant $\delta$ is
controlled by temperature, low (high) values of
$\ell_{\mbox{\scriptsize SED}}$ should be realized at low (high)
$T$. Hence, the observation of $\beta \approx 0.25$ at low $T$ and
$\beta \approx 0.5$ at high $T$ in \cite{eff94} could be due to the
influence of SED. In this case, the change of the scaling exponent
should be reflected by a change of the morphology.  

We find slope selection (hence $\alpha = 1$) for all simulations with
non--zero step edge diffusion. With $\ell_{\mbox{\scriptsize SED}}=1$
we find $\beta = 0.25$ resp.~$z=4.0$. With increasing step edge
diffusion the growth exponent rises continuously to $\beta =
0.33$. For intermediate values of $2 R_i \ll \ell_{\mbox{\scriptsize
SED}} \ll L$ we observe initially the growth exponent of SED--assisted
coarsening $\beta = 0.45$. When the size of the structures becomes
comparable to the step edge diffusion length, $\beta$ changes
to $\beta = 0.33$. Preliminary results for finite values of the
Ehrlich--Schwoebel barrier with our model (not discussed here) as well as
kinetic Monte--Carlo simulations \cite{ssb99} indicate that the
coarsening behavior is not affected by the actual value of the
Ehrlich--Schwoebel barrier. Clearly, small Ehrlich--Schwoebel barriers
increase the time for mound formation and the asymptotic scaling
exponents will be more difficult to obtain via computer simulations.

Assuming mass currents driven by surface energetics, the authors of
\cite{tsv97} obtain also $z=4$ in $d=2$ dimensions. Even though SED
might be interpreted as such a mass current, we obtain a significantly
different result for $\ell_{\mbox{\scriptsize SED}} \ge 4$. The reason
for this discrepancy remains unclear at present.

Forthcoming investigations will address more realistic models, for
instance with finite energy barriers at downward steps for which we
expect more insight into the increase of the growth exponent from
$\beta = 0.25$ to $\beta = 0.33$. For additional barriers in step
edge diffusion at corners we expect a non-trivial crossover behavior,
as well.

\begin{ack}
This work is supported by the Deutsche Forschungsgemeinschaft
through SFB 410.
\end{ack}

\bibliography{/users1/schinzer/Arbeit/Preprints,/users1/schinzer/Arbeit/Literatur,/users1/schinzer/Arbeit/NeueLit}
\bibliographystyle{unsrt}

\end{document}